# Optically probing the asymmetric interlayer coupling in rhombohedral-stacked MoS$_2$ bilayer

Jing Liang[1,2,#], Dongyang Yang[1,2,#], Jingda Wu[1,2,#], Jerry I Dadap[1,2], Kenji Watanabe[3], Takashi Taniguchi[4], Ziliang Ye[1,2]*

[1] Quantum Matter Institute, The University of British Columbia, Vancouver, BC V6T 1Z4, Canada

[2] Department of Physics and Astronomy, The University of British Columbia, Vancouver, BC V6T 1Z1, Canada

[3] Research Center for Functional Materials, National Institute for Materials Science, 1-1 Namiki, Tsukuba 305-0044, Japan

[4] International Center for Materials Nanoarchitectonics, National Institute for Materials Science, 1-1 Namiki, Tsukuba 305-0044, Japan

# These authors contributed equally to this manuscript.

* Correspondence: zlye@phas.ubc.ca

**Abstract**

The interlayer coupling is emerging as a new parameter for tuning the physical properties of two-dimensional (2D) van der Waals materials. When two identical semiconductor monolayers are stacked with a twist angle, the periodic interlayer coupling modulation due to the moiré superlattice may endow exotic physical phenomena, such as moiré excitons and correlated electronic phases. To gain insight into these new phenomena, it is crucial to unveil the underlying coupling between atomic layers. Recently, the rhombohedral-stacked transition metal dichalcogenide (TMD) bilayer has attracted significant interest because of the emergence of an out-of-plane polarization from non-ferroelectric monolayer constituents. However, as a key parameter responsible for the physical properties, the interlayer coupling and its relationship with ferroelectricity in them remain elusive. Here we probe the asymmetric interlayer coupling between the conduction band of one layer and the valence band from the other layer in a 3R-$MoS_2$ bilayer, which can be understood as a result of a layer-dependent Berry phase winding. By performing optical spectroscopy in a dual-gated device, we show a type-II band alignment exists at K points in the 3R-$MoS_2$ bilayer. Furthermore, by unraveling various contributions to the band offset, we quantitatively determine the asymmetric interlayer coupling and spontaneous polarization in 3R-$MoS_2$. Our results unveil the physical nature of stacking-induced ferroelectricity in TMD homostructures and have important implications for many-body states in the moiré superlattice of 2D semiconductors.

**Main**

Interlayer coupling, a ubiquitous ingredient in van der Waals (vdW) materials, offers unprecedented freedom to tailor the electronic band structure of two-dimension (2D) materials. Through precise control of the layer orientation in an artificial 2D assembly, one can create a periodically modulated interlayer coupling in moiré superlattice, which leads to many intriguing phenomena different from their monolayer substituents[1-6]. In semiconducting transition metal dichalcogenides (TMDs), such moiré potentials can quench the kinetic energy of electrons and localize them at the energy local minimum of the moiré superlattice, giving rise to a new platform for creating quantum emitter arrays and simulating various correlated physics[7-17]. In order to have a full grasp over these emerging phenomena, it is therefore important to understand the underlying interlayer coupling at different localization sites with different layer orientations.

Recently, parallel stacked TMD layers have attracted significant attention as it has a broken mirror symmetry with an out-of-plane spontaneous electrical polarization[18-25]. In the artificially stacked TMD bilayer with a marginal twist angle, the crystal structure spontaneously relaxes into multiple rhombohedral domains with alternating polarization[26], which can be electrically switched via in-plane sliding motion. As a result, the rhombohedrally (R) stacked TMD becomes a new ferroelectric semiconductor with a promising range of electronic and optoelectronic applications[24,25,27]. Here we optically probe the spontaneous polarization and interlayer coupling in an R-stacked TMD bilayer. Unlike in the common hexagonal polytype, the interlayer coupling in the rhombohedral polytype is interestingly asymmetric: The conduction band in one layer only couples to the valence band in the other, but not vice versa, due to the layer-dependent Berry phase effect. Such an asymmetric coupling is also the electric origin of spontaneous polarization and interlayer potential in TMDs with natural rhombohedral crystal structure or artificial parallel stacking with a marginal twist. We carry out our measurements by performing both electric-field and doping dependent optical spectroscopy in a dual-gated device made of a homogeneous bilayer $MoS_2$ exfoliated from a chemically synthesized 3R crystal.

In a 3R-MoS₂ bilayer, the adjacent layers are stacked in the same direction (Fig. 1a). A relative interlayer displacement along armchair direction gives two inequivalent stacking configurations: BA or AB stacking order, where the S atom in the top layer lies directly on the top of the Mo atom in the bottom layer for BA stacking order[28,29]. The parallel stacking direction maintains the non-inversion symmetry of the monolayer while the lateral shift breaks the mirror symmetry between the top and bottom layer, resulting in the $C_{3v}$ point group of the 3R-MoS₂ bilayer. Under $C_3$ rotational operation, the phase of the Bloch states at K points, the high symmetry points at the Brillion zone edge, has two contributions, $C_3\psi_K = e^{-i(m+m')\frac{2\pi}{3}}\psi_K$, where $m$ is the magnetic quantum number associated with the rotation of the atomic orbital around itself and $m'$ is associated with the Berry phase when hopping from one site to the next, which is dependent on the rotation center (Fig. 1b)[30-34]. The sum $m + m'$ is the total azimuthal quantum number (AQN).

The dependence of the Berry phase on the rotation center has a profound impact on interlayer coupling in 3R-MoS₂. In the monolayer, the AQNs between the conduction and valence bands are different by $+1$ or $-1$, depending on the valley index, which gives rise to the well-known valley selection rule[30]. In the 3R-MoS₂ bilayer, the lateral shift leads to a distinct rotational center of two layers, which adds an unequal Berry phase to different layers (Fig. 1b). Take the BA-stacked 3R-MoS₂ bilayer as an example: The states at K points in the bottom layer obtain an extra quantum number from the Berry phase of $m' = +1$, resulting in the overall AQN of $+1$ and $0$ for the conduction and valence bands (Fig. 1c). In contrast, the AQN of the conduction and valence bands in the top layer remains $0$ and $-1$. Since electrons can only tunnel between bands of the same AQN, the interlayer coupling $t$ only exists between the conduction band of the top layer and the valence band of the bottom layer at K points. Neither the conduction nor the valence bands at K points are directly coupled between the two layers, so the band edges can be distinctively defined for each layer in the 3R-MoS₂ bilayer at K points, in contrast to the strongly hybridized valence band at Γ point. Such an interesting asymmetric interlayer coupling induces a level repulsion between the two coupled bands and splits the layer degeneracy, resulting in a staggered gap at K points where the valence band edge at the Brillouin zone corner is localized in the top layer while the conduction band edge is localized in the bottom layer, which we refer to as an effective type-II band alignment at K points in a 3R-MoS$_2$ homobilayer (Fig. 1c)[27,33-35].

On the other hand, the asymmetric interlayer coupling also mixes the Bloch wave functions from different layers, which causes the Wannier center of the valence band from the bottom layer to shift towards the top layer while the valence band of the top layer is unchanged. Consequently, the overall Wannier center of all occupied valence band states shifts to the top layer, resulting in a downward out-of-plane polarization ($P$) according to modern Berry phase theory (Fig. 1d)[33,34,36-38]. The interlayer electrostatic potential associated with the polarization, $\phi_0 = Pd_0/\varepsilon_0\varepsilon_m$, generates an additional band offset in the 3R-MoS$_2$ bilayer, where $\varepsilon_0$ and $\varepsilon_m$ denote the vacuum permittivity and the out-of-plane dielectric constant of MoS$_2$ respectively, and $d_0$ is the interlayer distance.

Last but not least, the local chemical environments of the two layers are nonequivalent due to the lateral shift. In the BA stacking, the Mo atom in the bottom layer lies directly beneath the S atom in the top layer but the Mo atom in the top layer does not coincide with the S atom in the bottom layer. Such an asymmetric atomic configuration is known for causing some high-order electronic coupling, which leads to a larger bandgap in the bottom layer than in the top[33,35]. The bandgap energy difference, $\delta$, has been observed in both artificial and natural TMD bilayers through optical spectroscopy (Fig. 1e). As a result, the total band offset at the K points in a 3R-MoS$_2$ bilayer has three contributions: the asymmetric interlayer coupling strength $t$, the polarization-induced intrinsic interlayer potential $\phi_0$, and the bandgap difference $\delta$, which, according to our effective four-band model can be approximately expressed as (Supplementary Fig. S1 and Note 1)

$$\Delta_c = e\phi + \frac{t^2}{E_g^d} - \frac{\delta}{2} \quad (1)$$

$$\Delta_v = \Delta_c + \delta = e\phi + \frac{t^2}{E_g^d} + \frac{\delta}{2} \quad (2)$$

Here $\Delta_c$ and $\Delta_v$ are the conduction and valence band offsets and $E_g^d$ denotes the direct bandgap of the top layer MoS$_2$. Clearly, if we measure the total band offsets and unravel the contributions from each term, the coupling constant $t$ can be determined.

In our experiment, we determine each contribution using doping and field-dependent optical spectroscopy. A bilayer MoS$_2$ exfoliated from bulk 3R crystal (HQ graphene Inc.) is employed in this work. In contrast to artificial stacks, the sample is uniformly stacked with a BA stacking order, which is assigned by the interlayer exciton dipole direction. A dual-gated device was fabricated from exfoliated vdW materials using a layer-by-layer dry-transfer method[39], which allows independent control of the doping density (*n*) and the vertical electric field ($E_z$) between two layers (Methods). Figure S2 illustrates the schematic device structure and the optical image of the dual-gated device employed in this study. According to the thickness ratio of the bottom hexagonal boron nitride (hBN) gate and top hBN gate, one can introduce the external doping or electric field by applying a top gate ($V_t$) in proportion to the bottom gate ($V_b$) voltage with the same or opposite polarity (Supplementary Note 2).

We first investigate the optical response of the 3R-MoS$_2$ bilayer as a function of carrier density by reflectance contrast (RC) spectroscopy (Methods). All measurements are performed at 8 K. We also take the energy derivative of the RC to highlight small features (Fig. 2b). As shown in Figure 2a, the doping-dependent RC spectrum can be divided into three regions. In region-II, the Fermi level is inside the bandgap of both layers. As in the case of its monolayer counterpart, the 3R-MoS$_2$ bilayer exhibits two prominent exciton peaks, termed A and B excitons associated with the direct optical transition at the K points[40]. In contrast to the monolayer, the intralayer A exciton is split into two peaks, the low-energy A-exciton at 1.918 eV ($X_{At}$) and the high-energy A-exciton at 1.929 eV ($X_{Ab}$). Such a peak splitting ($\delta$ = 11.0±0.5 meV) has been observed in the artificial and natural TMD bilayers and has been attributed to the intralayer exciton emission from two layers, which is a powerful tool for probing the optical response from different layers. Here the photon energy difference of the intralayer excitons is directly read from the peak separation in the RC spectra since the gate-dependent RC spectra are complex due to the hBN encapsulation and the presence of top and bottom gates. The uncertainty of the bandgap difference is determined by the spectral resolution of 0.5 meV. Furthermore, the full width at half maximum of the intralayer exciton in the 3R-MoS$_2$ bilayer is about 12 meV, broader than that of the monolayer MoS$_2$[41], which could arise from the phonon scattering-induced exciton lifetime reduction as the MoS$_2$ bilayer is an indirect-gap semiconductor with a valence band maximum at the Γ point.

In the electron doping region-I, the electron doping dependences of the two A-excitons $X_{At}$ and $X_{Ab}$ are clearly different (Fig. 2a, b). With increasing electron doping density, the high-energy $X_{Ab}$ blueshifts and transfers its oscillator strength to another emergent peak on the low energy side of $X_{At}$. The observation can be explained by the interaction between the exciton and degenerate Fermi sea (FS) of excess charge carriers. As shown in Figure 2c, the $X_{Ab}$ excitons are dressed by excitations (electron-hole pairs) of the FS and split into two branches: a low-energy attractive exciton-polaron ($XP_{Ab}^{-\prime}$) and a high-energy repulsive exciton-polaron ($XP_{Ab}^{-}$)[42-46]. Around the transition area of region-I and region-II, we estimate the binding energy of $XP_{Ab}^{-\prime}$ to be 25 meV. With increasing electrostatic gating, the exciton-FS interaction is enhanced due to the expanding FS, which enlarges the repulsion between the $XP_{Ab}^{-}$ and $XP_{Ab}^{-\prime}$. Consequently, the $XP_{Ab}^{-\prime}$ redshifts whereas the $XP_{Ab}^{-}$ blueshifts with increasing electron density, and the high-energy repulsive exciton-polaron branch $XP_{Ab}^{-}$ transfers its oscillator strength to the low-energy attractive exciton-polaron branch $XP_{Ab}^{-\prime}$, which is consistent with previous studies of monolayer MoS$_2$[40]. On the other hand, the intralayer A-exciton from the top layer $X_{At}$ remains largely unchanged except for a slight initial redshift, potentially due to the weak screening or a Pauli-blocking-free exciton polaron effect with the charges in the other layer (Fig. 2c). The different doping dependence of the two A-excitons suggests that the electrons are doped into the layer with the larger bandgap, indicating a type-II band alignment at the K points of the 3R-MoS$_2$ bilayer[47].

In contrast to the distinctive response in electron doping, the two MoS$_2$ layers respond similarly to hole doping. The two prominent exciton peaks $X_{At}$ and $X_{Ab}$ are quickly replaced by two emergent peaks with lower energy, which redshift together with the increasing negative voltage (Figure 2a, b). Such a doping dependence arises from the valence band maximum at the Γ point in the 3R-MoS$_2$ bilayer, where the two layers hybridize strongly. When holes are doped into the bilayer, they become equally distributed between two layers. As a result, both $X_{At}$ and $X_{Ab}$ excitons are dressed by electron-hole pairs of the FS at the Γ point and become split into low-energy attractive exciton-polarons ($XP_{At}^{+\prime}$ and $XP_{Ab}^{+\prime}$) and high-energy repulsive exciton-polarons ($XP_{At}^{+}$ and $XP_{Ab}^{+}$), as shown in the first energy derivative of the RC spectra (Fig. 2b, c). The more abrupt oscillator strength evolution than that in the electron-doping region might originate from the higher density of states at the Γ point associated with the larger effective mass. Since $XP_{At}^{+}$ is expected to have similar energy as $XP_{Ab}^{+\prime}$ at low doping and should rapidly diminish with increasing doping, we cannot distinguish them in the first derivative spectra, but their opposite doping dependence can be found in the second derivative spectra (Fig. S3). As hole doping density further increases, the FS expands and consequently, the $XP_{At}^{+\prime}$ and $XP_{Ab}^{+\prime}$ redshift at the same rate. The binding energy of $XP_{At}^{+\prime}$ ($XP_{Ab}^{+\prime}$) at a low doping density limit is estimated to be 18 meV (11 meV). The binding energy difference could be caused by the finite layer polarization of the hole at the Γ point due to the interlayer potential[35]. Our doping assignment also agrees with the photoluminescence (PL) spectra in Figure S4. The doping dependence of optical response could also be explained by the trion model without affecting the main conclusion of this work[40,47-49].

Next, we utilize the field dependence of the RC spectra in the electron-doping region-I to determine the intrinsic interlayer potential $\phi_0$. Since the exciton is mainly dressed with the electron doped in the same layer to form exciton polarons, as shown in Figure 2, the oscillator strengths of the excitons and exciton-polarons become efficient probes of the doping imbalance between two layers, which is determined by the conduction band offset $\Delta_c$. As a part of $\Delta_c$ arises from the interlayer potential associated with the electric field between two layers, one can tune $\Delta_c$ by applying an external electric field with an antisymmetric top and bottom gate, $\Delta V = V_t - V_b$. When $\Delta_c$ becomes zero, the doped electrons are equally distributed in two layers, and $X_{At}$ and $X_{Ab}$ should have the same oscillator strengths. The specific relationship between $\Delta_c$ and $\Delta V$ depends on the ratio between the averaged gate capacitance $\bar{C}$ and the geometric and quantum capacitance of MoS₂ ($C_m$ and $C_q$), which can be calculated from the film thickness and dielectric constants (Supplementary Note 2, Figure S5). In particular, when the Fermi level is in the conduction band of both layers

$$\Delta_c \approx \frac{2eC_m}{2C_m+C_q+2\bar{C}}\phi_0 - \frac{2e\bar{C}}{2C_m+C_q+2\bar{C}}\Delta V \qquad (3)$$

where $e$ is the electron charge and $\phi_0$ is the intrinsic interlayer potential induced by the depolarization field. We note $\Delta_c$ has a different $\phi_0$ dependence when the Fermi level is inside the gap of one layer or two. Only when the Fermi level is in the conduction bands of both layers, the intrinsic interlayer potential $\phi_0$ can be obtained by measuring $\Delta V$ at $\Delta_c = 0$ condition (Supplementary Note 2).

We now consider the electric field dependence of the RC spectra at a fixed electron doping density ($n = 6 \times 10^{11}$ cm$^{-2}$). As $V_t$ is swept between 5.5 and 9.5 V, $V_b$ is always 0.85 times smaller than $V_t$ with an opposite polarity to compensate for the hBN thickness difference (Fig. 3a). When the external field is small ($V_t <$ 6.5 V), the electrons are located in the bottom layer. So the exciton in the top layer (X$_{At}$) is observed together with the exciton-polaron in the bottom layer (XP$_{Ab}^{-}{}'$ and XP$_{Ab}^{-}$), similar to region-I in Figure 2a. With increasing electric field, the offset between two conduction bands decreases, until the Fermi level reaches the conduction band edge of the top layer ($V_t =$ 6.5 V). From this point on, the electron begins to migrate from the bottom to the top layer (Fig. 3d). The interaction between X$_{At}$ and FS in the top layer forms an attractive and repulsive exciton-polaron, XP$_{At}^{-}{}'$ and XP$_{At}^{-}$. As XP$_{At}^{-}$ blueshifts and diminishes with the expanding FS, the XP$_{Ab}^{-}$ redshifts, and gains oscillator strength. Such an oscillator strength transfer terminates when the Fermi level is away sufficiently from the conduction band edge of the bottom layer ($V_t =$ 8.2 V).

As discussed above, we can quantify the intrinsic interlayer potential $\phi_0$ by measuring when the electrons become equally distributed in two layers (middle panel in Figure 3d). Since the oscillator strength of the intralayer exciton is a sensitive indicator of the electron doping in that layer, we extract the strength of their peaks from the RC spectrum at fixed photon energy (1.917 and 1.928 eV). Because the two excitons can have different intrinsic oscillator strengths and their observed peak strength can be convoluted with the local field factor and the exciton-polaron formation in the neighbouring layer, we normalize the peak strength change with respect to the maximum value within our gate tuning range (Figure 3e). Unsurprisingly, we observe a slope in both excitons where the peak strength varies rapidly with the gate voltage. The middle point of the slope region coincides with the crossing point of the two curves ($V_t =$ 7.35$\pm$0.15 V) indicating the $\Delta_c=$ 0 condition. The uncertainty is estimated from the difference between two middle points and the crossing point is found to be largely insensitive to the total doping density $n$. According to equation (3), the intrinsic interlayer potential $\phi_0$ is 58±1.5 mV, corresponding to an out-of-plane polarization, $P$ =0.55±0.02 µC/cm$^2$. Our measured interlayer potential is about 20% larger than the previous report in artificial stacks[24,25], which is expected since our optical measurement is performed in a homogenous sample without mixed domains and does not require domain flipping. Additionally, the downward polarization also confirms the BA stacking-order assignment (Supplementary Note 2). Multiple sets of field-dependent measurements are taken at different locations, and they all show quantitatively similar results as Figure 3a, confirming the homogeneous BA-stacking order of our sample and the intrinsic nature of our observation. An example of a field-dependent measurement at a different location is shown in Figure S6. The determination of charge distribution via exciton contrast is reliable since the low exciton density excited by the broadband white light source during the RC measurement does not affect the carrier distribution and the finite doped electrons do not cause significant screening of excitons.

To quantify the asymmetric interlayer coupling strength $t$, we also need to know the total intrinsic band offset. Here we measure this offset optically by investigating the photoluminescence (PL) spectra of the interlayer excitons. It has been reported that in the bilayer MoSe$_2$, the interlayer exciton comprises a hole residing in the Γ point and an electron at the K point (Fig. 4b)[35]. Since the electrons located in the layers with different band edges emit photons of different wavelengths, the intrinsic conduction band offset $\Delta_c$ can be measured by resolving the interlayer exciton PL peaks. In our dual gated 3R-MoS$_2$ bilayer device, the intrinsic interlayer PL spectrum has six peaks around 1.45 eV. These six peaks can be grouped into three pairs with an energy separation of about 20 meV in each pair. One prior explanation for such an energy separation is that each interlayer exciton has a phonon replica since phonons are needed for momentum-indirect optical transitions[50]. Available phonons for scattering the electron from the K to Γ point have energies of ~26 meV for the acoustic branch, and ~47 meV for the optical branch[50]. The energy splitting in each pair of interlayer excitons is ~20 meV, comparable to the energy difference between acoustic and optical phonons, suggesting the two exciton peaks can be assisted by different phonon branches. More studies are required to determine the specific phonon modes responsible for the optical transition. In the following, we focus on the low-energy branch of each pair.

The origin of the three interlayer exciton pairs is clarified by measuring their doping and electric field dependence. Since the two high-energy pairs disappear in the doped regimes (Fig. 4a), we attribute them to the intrinsic interlayer exciton, $X_{Ib}$ and $X_{It}$. The highest-energy pair is emitted from the top layer as the top layer has a higher conduction band than the bottom one at zero field. The broad peaks of the lowest energy are likely to result from trions emitted from the bottom layer ($T_{It}$). They are observable in region-II since the bilayer is always doped by bond charges induced by the spontaneous polarization. Such an assignment is confirmed by the electric field dependence of the emission energy (Fig. 4c). With $E_z$ varying from negative to positive, $X_{It}$ redshifts and $X_{Ib}$ blueshifts, respectively, and they cross each other at $V_t =6.5$ V. Similar to the picture in Figure 3, such a Stark shift is due to the tuning of the conduction band offset at the K point, while the valence band maximum at Γ point remains largely unchanged since the two layers hybridize strongly. When the conduction band offset is zero ($\Delta_c = 0$), $X_{It}$ and $X_{Ib}$ have the same photon energy corresponding to the crossing point in Figure 4c. The dipole moment of the interlayer exciton can be extracted from the slope of the Stark shift to be 0.31 $e·$nm, which agrees with the picture that the electron is localized in one layer while the hole is shared between the two, so their out-of-plane distance is about half of the interlayer distance ($d$~0.65 nm). The measured dipole moment agrees with the calculated value of the Γ-K interlayer exciton, which is much larger than that of the Γ-Q interlayer exciton, thus further confirming our assignment[35]. Interestingly, we find the interlayer trion peaks have similar field-dependent Stark shift as the interlayer exciton.

When $E_z = 0$, the energy difference between $X_{It}$ and $X_{Ib}$ of 58.0±0.5 meV is equal to the intrinsic conduction band offset $\Delta_c$. The uncertainty of $\Delta_c$ is also determined by the spectral resolution of 0.5 meV as that of $\delta$. The spin configuration of conduction bands should not affect the determination as long as the two sets of spin bands have the same band offset. Considering the optically determined $\Delta_c$, $\delta$, and $\phi_0$, we extract the asymmetric interlayer coupling of the 3R-MoS$_2$ bilayer to be $t =100±25$ meV. This conclusion demonstrates that the asymmetric interlayer coupling is not negligible, in contrast to some estimation in the literature[38]. Instead, our measured asymmetric coupling strength in the 3R bilayer is of the same order as the predicted interlayer coupling strength between valence bands in the 2H bilayer[38], in agreement with first-principles calculations[34]. Our four-band Hamiltonian should be regarded as an effective model which describes the essential physics involving states near the conduction and valence band edges of the 3R-MoS$_2$ bilayer. As such, the interlayer coupling parameter $t$ should also be regarded as an effective parameter which includes, in principle, contributions from all interlayer tunneling processes between the conduction band in one layer and the valence band in the other. Besides the bandgap difference $\delta$, the higher order corrections of remote bands to the band offset are qualitatively discussed in Supplementary Note 1.

As a summary, the polarization-induced intrinsic interlayer potential $\phi_0$, the interlayer coupling $t$, and the direct bandgap difference $\delta$ correspondingly contribute 58±1.5 meV, 6±2.5 meV, and −5.5±0.25 meV to the intrinsic conduction band offset $\Delta_c$ of 58 meV at K points in 3R-MoS$_2$ (Table I). The experimental results agree with previous self-consistent calculations which relates the intrinsic interlayer coupling and interlayer potential[27]. Since $\phi_0$ is only determined by the ratio of the averaged gate capacitance $\bar{C}$ and the geometric capacitance of bilayer MoS$_2$ ($C_m$), neither the size of the spin-orbit coupling induced conduction band splitting nor the spin configuration of the lowest band affect our measurement conclusions[51]. The contribution from the Q points is not expected to have a qualitative impact on our conclusions either for the same reason.

In conclusion, we leverage the distinctive sensitivity of Fermi polaron to charges doped in different layers to probe the intrinsic interlayer potential, and thus the spontaneous polarization in a homogeneous 3R-MoS$_2$ bilayer. Compared to the graphene electrical sensing approach to determining the spontaneous polarization, our optical technique does not rely on polarization switching via interlayer sliding, rendering it suitable for quantifying the spontaneous polarization in a wide range of materials. In combination with the optically measured band gaps and band offsets, we quantitatively determine the strength of an asymmetric interlayer coupling at K points, which is fundamental to the ferroelectric and optoelectronic applications of semiconducting TMDs where sliding ferroelectricity has been observed[24]. Last but not least, our results lay an important foundation for understanding the moiré superlattice formed by hetero- or twisted homo-structures. Besides the exotic physics predicted from the position-dependent interlayer coupling[52], the hopping between layers is known for determining the moiré potential depth as well as the bandwidth of the moiré bands, whose competition with the Coulomb interaction has led to a range of correlated insulating states observed in various experiments[53]. Therefore, the full understanding of interlayer coupling in different stacking orders is crucial for exploring these new semiconducting moiré materials.

We acknowledge support from the Natural Sciences and Engineering Research Council of Canada, Canada Foundation for Innovation, New Frontiers in Research Fund, Canada First Research Excellence Fund, and Max Planck–UBC–UTokyo Centre for Quantum Materials. Z.Y. is also supported by the Canada Research Chairs Program. The authors would like to thank Benjamin T. Zhou for the helpful discussion.

**Figures and captions:**

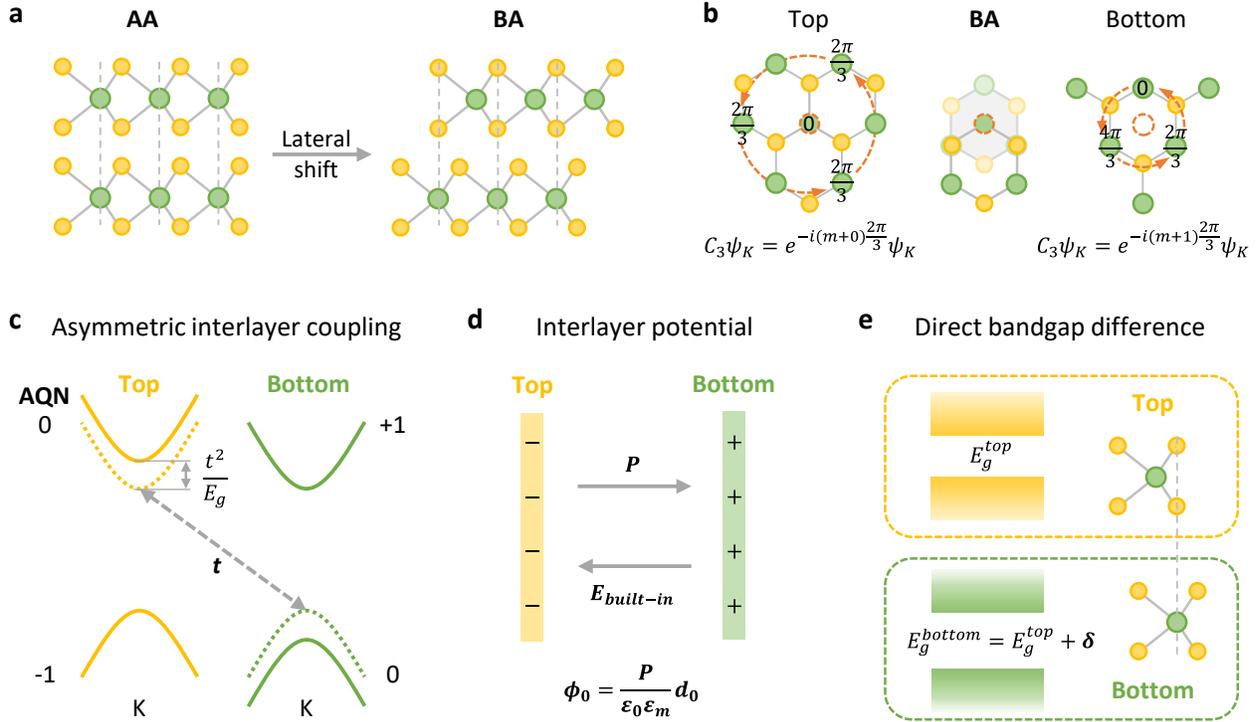

**Figure 1 | Schematic of asymmetric interlayer coupling in the 3R-MoS$_2$ bilayer.** (**a**) Side view of atomic structures of the 3R-MoS$_2$ bilayer with AA (left) and BA (right) stacking configurations, where the yellow and green spheres denote the S and Mo atoms, respectively. (**b**) Schematic of the phase winding at K points in monolayer MoS$_2$ when the $C_3$ rotational center is located at a Mo site (left panel) and the hollow center of the hexagon formed by Mo and S (right panel). The middle panel is the top view of the atomic structure of the 3R-MoS$_2$ bilayer with BA stacking configuration. The small, dashed orange circle denotes the $C_3$ rotational center. (**c**) Electronic band structure at K points of a 3R-MoS$_2$ bilayer with BA-stacking configuration. Yellow (green) dashed lines denote the original uncoupled bands. Yellow (green) solid lines denote the bands of the top (bottom) MoS$_2$ layer. Numbers denote the overall azimuthal quantum number (AQN) of the conduction/valence band edge at K points. Interlayer coupling $t$ only exists between the conduction band of the top layer and the valence band of the bottom layer, resulting in a level repulsion $t^2/E_g$ between these two bands, where $E_g$ is the direct bandgap of MoS$_2$. (**d**) Schematic of spontaneous out-of-plane

polarization *P* pointing from the top layer (yellow) to the bottom layer (green) in a 3R-MoS$_2$ bilayer with BA-stacking configuration. **(e)** Schematic of direct bandgap at the K point of the top (yellow) and bottom (green) layer in a 3R-MoS$_2$ bilayer with BA-stacking configuration. The atomic structure shows the nonequivalent environments of the Mo atom in the top and bottom MoS$_2$ layers.

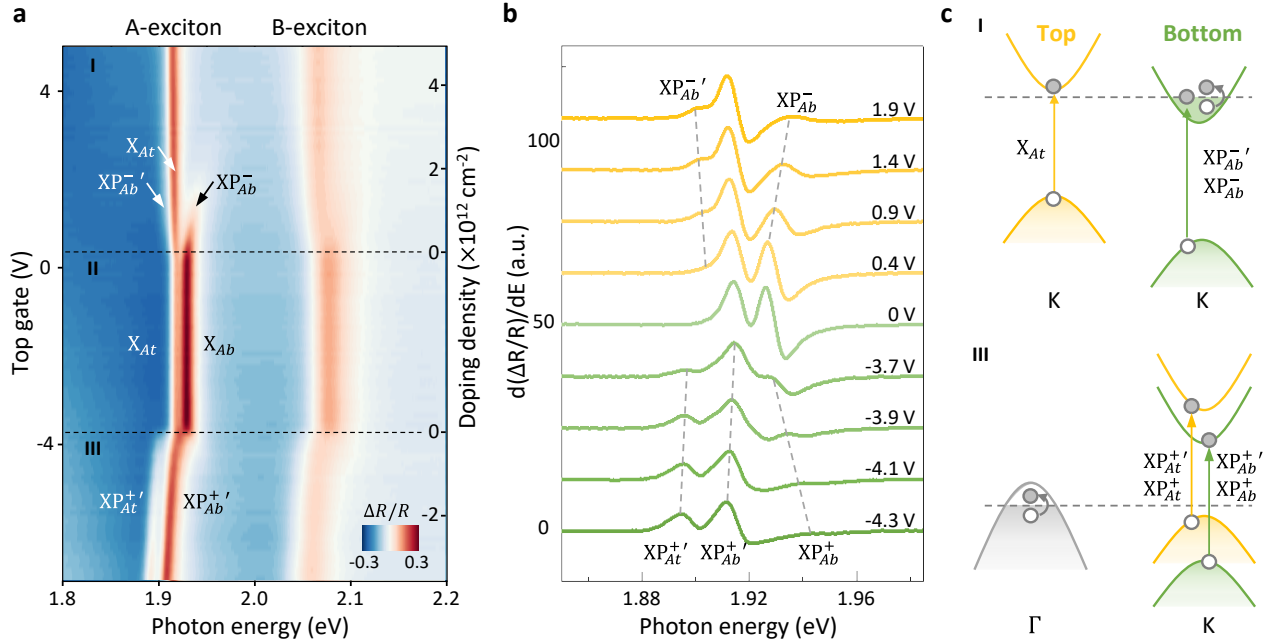

**Figure 2 | Doping-dependent reflectance contrast spectrum of intralayer excitons in the 3R-MoS$_2$ bilayer.** (**a**) Contour plot of the doping-dependent reflectance contrast spectra of the 3R-MoS$_2$ bilayer. The gate voltage is applied proportionally on top and bottom gates with relation $V_b = 0.85 V_t$. The right y-axis denotes the corresponding doping density (positive *n* for electron doping and negative *n* for hole doping). The black dashed lines divide the spectrum into three regions when the Fermi level is in the (I) conduction band, (II) bandgap, and (III) valence band. (**b**) First energy derivative of the reflectance contrast spectrum at different top gate voltage. Dashed gray lines denote different exciton-polaron in region-I and region-III. (**c**) The electronic band structure of the 3R-MoS$_2$ bilayer with BA stacking configuration at the region-I (top panel) or region III (bottom panel) in (a). The yellow (green) color denotes the electronic states localized in the top layer (bottom layer). The dashed gray line denotes the Fermi level. In region I, electrons are doped into the conduction band at the K point of the bottom layer. $XP_{Ab}^{-\prime}$ and $XP_{Ab}^{-}$ represent attractive and repulsive exciton-polarons, respectively, from the bottom layer when the intralayer excitons $X_{Ab}$ are dressed with electron-hole pairs in the Fermi sea (FS). In region III, holes are doped into Γ point. $XP_{Ai}^{+\prime}$ and $XP_{Ai}^{+}$ represent attractive and repulsive exciton-polarons, respectively. The subscript *i* is *t* or *b*, which denotes the optical transitions at the K points from the top or bottom layers.

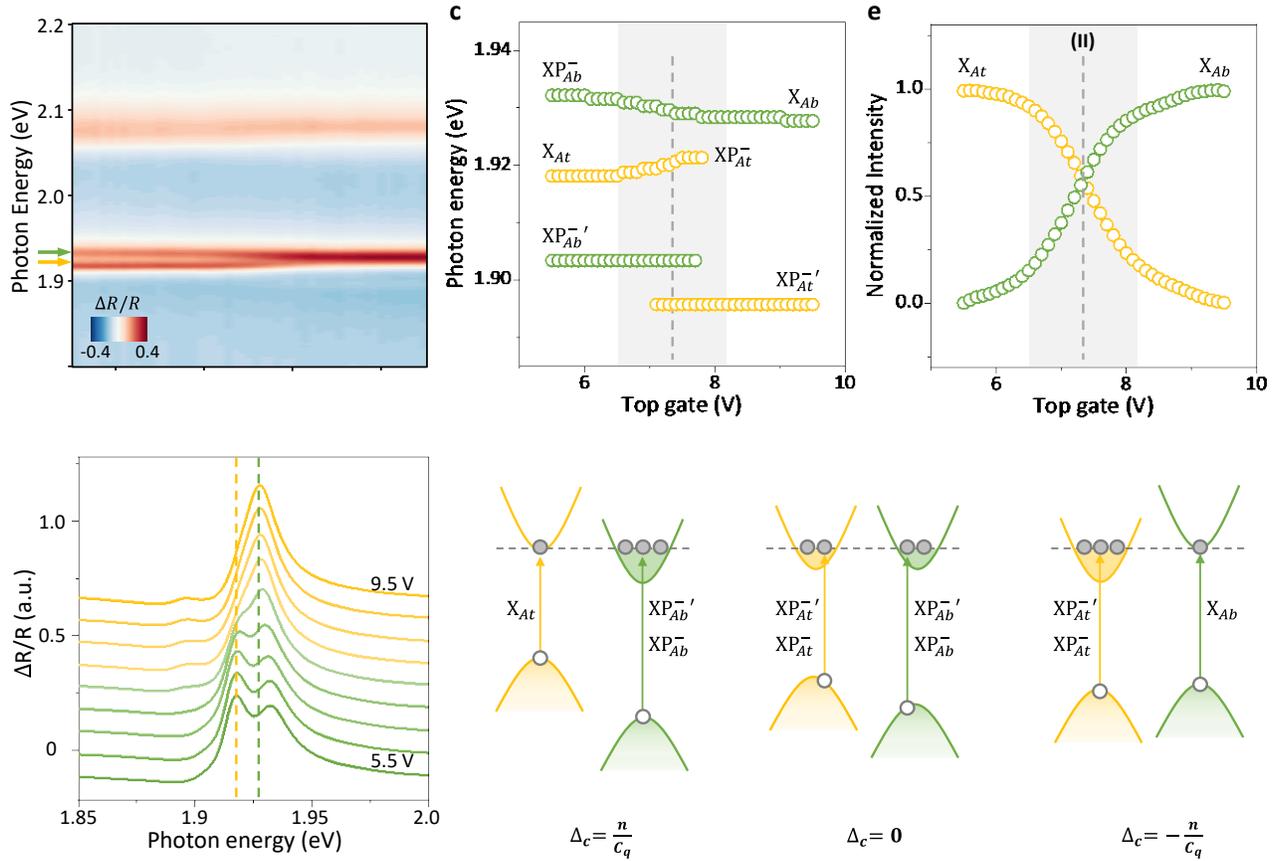

**Figure 3 | Electric-field-dependent reflectance contrast spectrum of intralayer excitons in 3R-MoS$_2$ bilayer at a fixed electron doping density.** (**a**) Contour plot of the electric-field-dependent reflectance contrast spectrum of 3R-MoS$_2$ bilayer at a fixed electron doping density. The yellow and green arrows denote the intralayer A-exciton, X$_{At}$ and X$_{Ab}$, from the top and bottom layers respectively. (**b**) Reflectance contrast spectrum at different top gate voltages in steps of 0.5 V. The bottom gate voltage is $V_b = -0.85 V_t + 1.65$. The yellow and green dashed lines denote X$_{At}$ and X$_{Ab}$ from the top and bottom layers, respectively. (**c**) From low-energy to high-energy, electric-field dependence of XP$_{At}^{-\prime}$ (attractive exciton-polaron in the top layer), XP$_{Ab}^{-\prime}$ (attractive exciton-polaron in the bottom layer), X$_{At}$ or XP$_{At}^-$ (intralayer exciton or repulsive exciton-polaron in the top layer) and X$_{Ab}$ or XP$_{Ab}^-$ (intralayer exciton or repulsive exciton-polaron in the bottom layer) energies extracted from (a). (**d**) Schematics of band alignment and existing optical transitions of a BA-stacked 3R-MoS$_2$ bilayer at a fixed electron doping density when (I) the Fermi level reaches the conduction

band edge of the top layer with $\Delta_c = n/C_q$, (II) the Fermi level is in the conduction band of both layers and electrons are equally doped into two layers with $\Delta_c = 0$, (III) the Fermi level is at the conduction band edge of the bottom layer with $\Delta_c = -n/C_q$. **(e)** Normalized intensity changes of $X_{At}$ and $X_{Ab}$ extracted from (a) along the fixed photon energy indicated by yellow and green arrows. The crossing point denotes the same oscillator strength of $XP_{At}^-$ and $XP_{Ab}^-$ when the conduction band offset is zero as shown in the middle panel of (d).

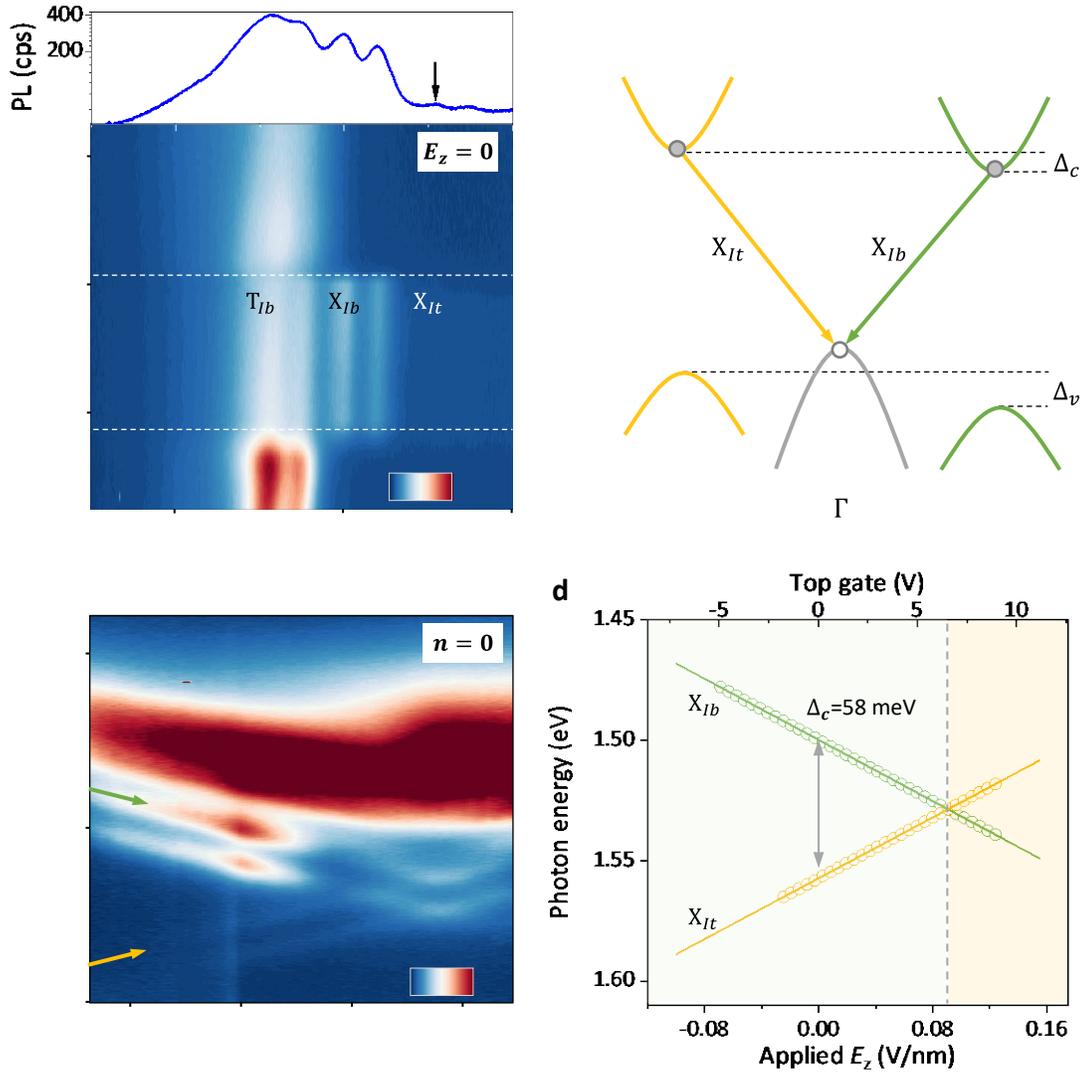

**Figure 4 | Doping-dependent and electric-field-dependent photoluminescence spectrum of Γ-K interlayer excitons in the 3R-MoS$_2$ bilayer.** (**a**) Contour plot of the doping-dependent PL spectrum of momentum-indirect interlayer excitons in the 3R-MoS$_2$ bilayer without external electric field. The gate voltage is applied proportionally on top and bottom gates with relation $V_b = 0.85V_t$. At zero doping, from low-energy to high-energy, three pairs of peaks indicate T$_{Ib}$ (interlayer trion), X$_{Ib}$ (interlayer Γ-K exciton), and X$_{It}$ (interlayer Γ-K exciton), respectively; X$_{It}$ is more evident in (c) as indicated by the yellow arrow. Top panel is an average PL spectrum in region-II with the black arrow denoting the X$_{It}$ (interlayer Γ-K exciton). cps: counts per second. (**b**) Type-II band alignment of BA-stacked MoS$_2$ bilayer at K points of the Brillouin zone and momentum-indirect Γ-K

transitions. $\Delta_c$ and $\Delta_v$ denote the conduction band offset and valence band offset at K points, respectively. **(c)** Contour plot of the electric-field dependent PL spectrum in the 3R-MoS$_2$ bilayer with zero doping. The green (yellow) arrow denotes the interlayer exciton $X_{Ib}$ ($X_{It}$). The gate voltage is applied proportionally on top and bottom gates with relation $V_b = -0.85V_t$. **(d)** Electric-field dependence of interlayer excitons extracted from (c). Solid lines denote the linear fitting results. The dashed gray line denotes the same energy of $X_{Ib}$ and $X_{It}$ when $\Delta_c=0$.

Table I | The measured conduction/valence band offset at the K points, the asymmetric interlayer coupling, the direct bandgap difference, and the polarization-induced interlayer potential in a 3R-MoS$_2$ bilayer.

| $\Delta_c$ | $\Delta_v$ | $t$ | $\delta$ | $\phi_0$ |
|---|---|---|---|---|
| 58 ± 0.5 meV | 69 ± 1 meV | 100 ± 25 meV | 11 ± 0.5 meV | 58 ± 1.5 mV |